\documentclass{emulateapj}
\shorttitle{Young multiplanet systems}
\shortauthors{Dodson-Robinson \& Salyk}
\makeatletter

\newcommand{\Rmnum}[1]{\expandafter\@slowromancap\romannumeral #1@}
\makeatother
\newcounter{minirefcount}
\newcommand{\miniref}[2]{\refstepcounter{minirefcount}\label{#2}(\arabic{minirefcount}) #1}

\begin{document}

\title{Transitional disks as signposts of young, multiplanet systems}

\author{Sarah E. Dodson-Robinson \& Colette Salyk}

\affil{Astronomy Department, University of Texas at Austin, 1 University
Station C1400, Austin, TX 78712 USA}

\email{sdr@astro.as.utexas.edu}

\begin{abstract}

Although there has yet been no undisputed discovery of a still-forming
planet embedded in a gaseous protoplanetary disk, the cleared inner
holes of transitional disks may be signposts of young planets. Here we
show that the subset of accreting transitional disks with wide,
optically thin inner holes of 15~AU or more can only be sculpted by
multiple planets orbiting inside each hole. Multiplanet systems provide
two key ingredients for explaining the origins of transitional disks.
First, multiple planets can clear wide inner holes where single planets
open only narrow gaps. Second, the confined, non-axisymmetric accretion
flows produced by multiple planets provide a way for an arbitrary amount
of mass transfer to occur through an apparently optically thin hole
without over-producing infrared excess flux. Rather than assuming the
gas and dust in the hole are evenly and axisymmetrically distributed,
one can construct an inner hole with apparently optically thin infrared
fluxes by covering a macroscopic fraction of the hole's surface area
with locally optically thick tidal tails. We also establish that other
clearing mechanisms, such as photoevaporation, cannot explain our subset
of accreting transitional disks with wide holes. Transitional disks are
therefore high-value targets for observational searches for young
planetary systems.

\end{abstract}


\section{Introduction}
\label{intro}

Protoplanetary disks of gas and dust surrounding young stars are the
birthplaces of planets---yet to date, there is no undisputed
observation of a still-forming planet in a gaseous disk. Astronomers
have therefore paid considerable attention to transitional
disks---protoplanetary disks with inner clearings---because such
clearings occur as the result of tidal interactions between disks and
planets (e.g.\ Lin \& Papaloizou 1993). To date, over 30 transitional
disks have been discovered with the IRS instrument aboard the {\it
Spitzer} Space Telescope (Muzerolle et al.\ 2010). If theorists were to
demonstrate a planetary origin for any inner hole in a transitional
disk, that disk would harbor extraordinary discovery potential for
young, still-forming planets. Indeed, gaps in debris disks surrounding
main-sequence stars harbor the giant planets Fomalhaut~b
\citep{kalas08}, $\epsilon$~Eridani~b \citep{hatzes00}, the HR~8799
system \citep{marois08, marois10} and $\beta$~Pictoris~b
\citep{lagrange10}.

Inner holes in gaseous protoplanetary disks with ages $\sim 1-10$~Myr
are initially identified by the flux deficits they leave in the near-IR
and/or mid-IR part of the disk's spectral energy distribution (SED). The
observed flux deficits correspond to ``missing'' dust with
characteristic temperatures of $\sim 200-1000$~K \citep[e.g][]{calvet02,
furlan06, brown07}. Model SEDs of the first transitional disks to be
discovered, GM~Aur \citep{koerner93, calvet05} and~TW Hya
\citep{calvet02}, were based on disks with small inner holes only 3-4~AU
wide. Soon after, investigations of planet-disk interactions
demonstrated that such clearings could naturally be produced by a single
Jupiter-mass planet \citep{quillen04, rice06}. Although the planet
opened only a narrow gap, not a hole extending all the way to the
central star, the dynamical studies showed that the residual gas and
dust inside the gap would accrete onto the central star on the viscous
timescale, widening the initial gap into a small, optically thin hole.

The first hole-size measurements performed with millimeter
interferometers showed remarkable consistency with SED-based models
\citep{brown09}.  However, further spatially resolved imaging of
transitional disks brought a surprise. Some disks for which small,
3-4~AU inner holes were inferred based on SED modeling---including the
canonical transitional disk, GM~Aur---were shown to have large inner
clearings of 15~AU or more once imaged at high angular resolution
\citep{hughes09, andrews10, andrews11}.  Yet, as the models of
\citet{rice06} and \citet{quillen04} hinted, each single planet can open
only a narrow gap, which must then widen into a hole by viscous clearing
of the material inside the gap. The extraordinarily long viscous
timescales at distances of 15~AU or more meant the simple theory that
each transitional disk hole contained a single giant planet was no
longer satisfactory.

The prosaic explanation for wide holes in transitional disks---tidal
truncation of the disk by a stellar companion, as observed in CoKu Tau/4
\citep{ireland08}---is also not likely responsible for the unusual and
complex disks that concern us in this article. Based on observations
with the Keck interferometer, \citet{pott10} rule out stellar companions
in 0.35-4 AU orbits at the 99.7\% confidence level in the holes of
DM~Tau, GM~Aur, LkCa~15, UX~Tau~A and RY~Tau---at least three of which
are examples of transitional disks with wide holes \citep{andrews11}.
\citet{kraus11} also found no stellar companions in the inner holes of
DM~Tau, GM~Aur, LkCa~15 and UX~Tau~A in their non-redundant aperture
masking survey of the Taurus star-forming region. The high accretion
rates in these sources (see Table \ref{table:stats} and references
therein), as well as the presence of inner disk dust and gas
\citep[e.g.][]{salyk09, espaillat10}, are also consistent with
sub-stellar rather than stellar  companions, as they indicate that disk
truncation is incomplete.  The set of transitional disks with large
inner holes, up to 75~AU, and significant accretion represents a
significant subset of transitional disks, and a subset that continues to
increase in number \citep{andrews11}. In Table \ref{table:stats}, we
list a sample of eight rapidly accreting transitional disks that have
holes between 15 AU and $> 70$~AU wide.

In general, transitional disks have a variety of observational
characteristics and may not represent a single physical phenomenon.  For
example, clearing driven by the magnetorotational instability
\citep{chiang07} can widen dynamically cleared holes once such holes
extend all the way to the dust-sublimation radius. In \S
\ref{otherclearing} we explain in detail why magnetorotational clearing
cannot be responsible for the particular set of transitional disks
listed in Table \ref{table:stats}.  Photoevaporation can also produce
disks with inner cleared regions. Prior work on UV photoevaporation
showed that evaporation-induced inner holes were inconsistent with
accretion rates higher than $\sim 10^{-10} \: M_\odot$~yr~$^{-1}$
\citep{alexander07}.  In essence, hole formation can only occur once
evaporation rates exceed accretion rates, as otherwise the hole will be
replenished by accretion.  More recently, however, it has been argued
that X-ray photoevaporation should be the dominant disk-clearing process
and produce much higher mass-loss rates than UV
\citep[e.g.][]{ercolano08}. Thus, the set of parameter space consistent
with photoevaporation has grown to include somewhat higher accretion
rates. It is entirely possible that photoevaporation produces some
transitional disks, while dynamical clearing by planets produces others.
However, even X-ray photoevaporation models do not appear to be able to
explain the existence of disks with {\it both} large holes $\gtrsim$
15--20 AU {\it and} significant accretion rates $\gtrsim
10^{-8.5}M_\odot$ yr$^{-1}$ \citep{owen11}.

There is yet one more possible explanation for accreting transitional
disks with wide holes---clearing by multiple planets, whose individual
gaps could overlap to form a wide hole. Since planets have gap-crossing
tidal tails and don't completely truncate the disk, a multiplanet system
would allow gas and dust to accrete through the gap. In this work, we
demonstrate that the particular set of transitional disks with wide
holes and high accretion rates listed in Table \ref{table:stats}
presents unambiguous evidence for tidal clearing by {\it multiple}
planets. While \citet{zhu11} found that wide, optically thin gaps in
accreting disks could not be produced by planets alone, their
conclusions were based on the assumption of axisymmetric gas and dust in
the hole. Instead, by taking into account the non-axisymmetric flow
patterns produced by planets, we argue not only that multiple planets
are consistent with high accretion rates and reduced infrared fluxes,
but that they represent the only possible explanation for our sample of
transitional disks.

We begin our argument for planetary influence on transitional disks in
\S \ref{diskproperties}, where we demonstrate that planetary systems
provide a viable explanation for the gas confinement, non-axisymmetry
and mid-IR flux deficits observed in transitional disks with large
holes. In \S \ref{oneplanet}, we show that the simplest planetary
configuration---a single gas giant inside the hole---cannot explain the
transitional disks with wide holes. We present dynamical models of
multiple planets opening overlapping gaps in \S \ref{multiplanets}. In
\S \ref{modelseds} we use the density distributions predicted
by the dynamical model in \S \ref{multiplanets} to produce SEDs, and 
discuss compatibility with observed photometry.   In \S \ref{photoevaporation} 
we delve into the photoevaporation mechanism in more detail, demonstrating conclusively
that it is not a viable originator of our sample of transitional disks.
Finally, we present conclusions and testable predictions in \S \ref{conclusions}.

\section{Properties of transitional disks with large holes}
\label{diskproperties}
The primary observational characteristic that defines all transitional
disks is that the near- and/or mid-IR fluxes are lower than those observed from
`typical' protoplanetary disks. Since the total flux
emitted from an optically thick disk is not affected by small changes in
the disk surface density, the deficit of flux in
transitional disks is generally interpreted as requiring depletion
sufficient to make the disk vertically optically thin at the
flux-deficient wavelengths. However, it is important to keep in mind
that the actual observable is the emitted {\it flux}, 
not the optical depth, $\tau_\nu$. Knowledge of
optical depth is required to convert observed fluxes into disk surface
densities, but the assumption that $\tau_\nu$ is uniform
throughout the hole, or even constant within a given annulus, can lead
to serious errors in estimating the mass distribution in the inner disk.
In particular, confined, optically thick `streamers' within a hole can produce the same
emergent flux as azimuthally symmetric, optically thin dust, while
hiding a significant amount of the dust and gas that must move through the
hole to match observed accretion rates. 
If the grains in a narrow tidal tail cover
more surface area than the tidal tail itself, so that the tail is
locally optically thick, some of the grain mass is effectively hidden.
Clumping in Saturn's rings produces exactly this type of plateau in
apparent optical depth as a function of particle surface density
(Robbins et al.\ 2010).



In the case of optically thick tidal tails surrounded by empty space,
one can simply replace $\tau_\nu$ with a geometrical ``filling
factor'', $f(r)$ --- the fraction of an annulus's
area covered by tidal tails --- in the standard equation for disk flux:
\begin{equation}
F_{\nu} = \int_{r_{\rm in}}^{r_{\rm out}} 2 \pi r dr B_{\nu}
\left [ T(r) \right ] f(r)/d^2,
\label{diskflux}
\end{equation}
where $F_{\nu}$ is the flux at a given frequency, $r_{\rm in}$ and
$r_{\rm out}$ are the disk inner and outer radii, respectively,
$B_{\nu}$ is the Planck function in frequency space, $d$ is the distance to the system, and $T$ is the dust
temperature at a given radius.
An important difference between $f$
and optical depth $\tau_\nu$, however, is that $f$ is purely geometric, and so independent of $\nu$,
while $\tau_\nu$ depends on the opacity $\kappa_\nu$, which is a function of $\nu$.  Therefore, the mass distribution and/or temperature
structure must be different in the two cases in order to produce the
same spectrum.  In addition, in a realistic disk, one must account for
the optically thin disk atmosphere, which produces silicate emission features
dependent on $\kappa_\nu$.

As an example of the ambiguities inherent in modeling transitional disk
SEDs, in Figure 1, we show two alternative inner disk dust distributions
that produce similar near-IR spectra. To model the resultant flux of the
models, we used the radiative transfer code RADMC \citep[similar to][but
without vertical structure integration]{dullemond04} but optically thin
components were also benchmarked with an analytical model. The first
model consists of an optically thin region from $0.22\footnote[1]{The
measured inner radius of GM~Aur's disk from near-IR interferometric
measurements \citep{akeson05}.}-1$ AU (with
$\Sigma_\mathrm{dust}=\Sigma_0 (R/\mathrm{AU})^{-0.6}$--- used for this
and all other models --- and $\Sigma_0=5\times10^{-8}$ g~cm$^{-2}$) and
a second optically thin region from 16--20 AU (with $\Sigma_0=10^{-7}$
g~cm$^{-2}$). The second model consists of an optically thin region from
$0.22-1$ AU (with $\Sigma_0=3\times10^{-8}$ g~cm$^{-2}$) and an
optically thick streamer from $0.7-13$ AU (with $f=0.03$ and
$H/R=0.08\times(R/\mathrm{AU})^{1/7}$). Both models use the same
standard Milky Way dust composition of 80\% small circumstellar
silicates and 20\% small carbon grains \citep{weingartner01, li01, draine07}, with
opacities from \citet{ossenkopf92} and \citet{zubko96} (``ACAR'').   The carbon grains, being
featureless, only act to reduce the line/continuum contrast, and so their quantity
is degenerate with other parameters, such as the inner disk small dust grain mass.
Optically thin surface densities were chosen to roughly match GM~Aur
photometric data (although an exhaustive analysis of parameter space was
not performed, and clearly these models have some trouble accurately
fitting the observed silicate feature).  The two models produce nearly
identical emergent spectra.  Figure 1 demonstrates that significant
radial mass transfer can occur through transitional disk holes and still
match photometric data, provided the mass is channeled into narrow
streams such as planetary tidal tails.

Besides extremely wide inner holes and high accretion rates, many
transitional disks contain additional signatures of dynamical
perturbation by planetary systems. \citet{brown09} noted significant
azimuthal asymmetries in the outer disk dust distribution for
LkH$\alpha$ 330 and HD 135344 B, while \citet{pontoppidan08} noted
asymmetries in the disks of HD 135344 B and SR 21 on the scale of a few AU or
less\footnote[2]{Although the asymmetries are observed in the
astrometric profile of a CO emission line, the asymmetry can be present
in either the gas or dust component.}.  Near-IR spectra were suggestive
of radial dust asymmetries such as eccentric inner holes in even early modeling studies
\citep[e.g.][]{calvet05}, and recent studies show that both
pre-transitional and transitional disks require multiple inner disk
components, including nearly evacuated regions, in order to match
observed fluxes and interferometric visibilities
\citep{espaillat10, andrews11, akeson11}.  Similarly, SR 21 has a
gas disk truncated at $\sim$7 AU \citep{pontoppidan08}, while the dust
disk is truncated at $\gtrsim 30$ AU \citep{brown09, andrews11}.
Finally, CO rovibrational lineshapes suggest that hot CO gas in
transitional disks may be dynamically confined to smaller regions than
in disks without inner holes (Salyk et al.\ 2011). The tidal barriers of
massive planets would provide an extremely effective way to limit the
radial distribution of hot gas in transitional disk holes.

A final, unusual characteristic of transitional disks is that they do
not seem to arise at any particular star age. Rather, \citet{furlan09}
show that the transitional disk fraction is 3-4\% in the Taurus,
Chamaeleon and Ophiuchus star-forming regions, despite an age spread of
1.5~Myr betweeen the clusters.  The transitional disk fractions reported
by Currie \& Sicilia-Aguilar (2011) also indicate that disks with inner
holes (as opposed to homologously depleted disks) are not preferentially
found in clusters of a particular age. The inner-hole configuration,
then, does not appear to represent an evolutionary stage all disks must
go through---in which case there would be a peak in the transitional
disk fraction for a given cluster age. \citet{furlan09} use their
observed transitional disk fractions to argue that the optically thin
inner-hole configuration lasts about 100,000 years, which is similar to
Type II planet migration timescales from the 10-15~AU region
\citep{ward97}. Note, however, the warning by Currie \& Sicilia-Aguilar
(2011) that there is a degeneracy between the time it takes a process to
operate, and the frequency with which it operates. It is possible that
$10^5$~years is the hole clearing timescale rather than the lifetime of
the inner hole.

Planets, therefore, have the potential to provide all the unusual
characteristics we observe in the subset of transitional disks
under investigation here (Table \ref{table:stats}):
\begin{itemize}
\item Wide, apparently optically thin inner holes with $r \ga 15$~AU,
\item Near and mid-IR flux decrements as compared to
classical protoplanetary disks,
\item High accretion rates of at least $\sim10^{-9} \; M_{\odot} \; {\rm
yr}^{-1}$,
\item Confined, non-axisymmetrically distributed gas and dust.
\end{itemize}
In the next section, we begin by examining the suitability
of the simplest planetary configuration, a single gas giant, for opening
the wide holes observed in our sample of transitional disks.


\section{On the feasibility of a single-planet origin for large
holes}
\label{oneplanet}

Clearing by a single giant planet is a commonly invoked explanation for
the gaps in transitional disks (e.g. Rice et al.\ 2003, Espaillat et
al.\ 2008, Furlan et al.\ 2009). In \S \ref{gapwidth} we quantify the
maximum gap width a planet can open, showing that it is not enough to
explain our sample of accreting transitional disks with wide holes. In
\S \ref{accretion} and \ref{otherclearing} we show that even when one
considers the combined effects of a gap-opening planet and accretion or
spiral wake-driven clearing of material inside the gap, the timescale
for opening a wide hole is outside the inferred $10^5$-year transitional
disk lifetime/clearing time.

\subsection{Maximum gap width}
\label{gapwidth}

To build a simple analytical framework for understanding the scope of
planetary influence on disks, we begin by imagining the disk as a swarm
of massless test particles. Treating the disk as if it were composed of
test particles on ballistic trajectories gives the maximum possible gap
width a planet can open because gas pressure and viscosity act to fill
in the dynamically cleared gap. We estimate the maximum gap width by
examining the time it takes for the planet to substantially alter the
angular momentum of close-passing particles. The planet opens a gap when
repeated encounters cause the particles at a given location to be
strongly scattered within the age of the star/disk system. We define
strong scattering as a change in the test particle's specific angular
momentum of a factor of 2-3. The analytical framework used here applies
to non-resonant particle orbits.

We consider a test particle with Keplerian frequency $\Omega$ and
unperturbed orbital radius $r_o$ during a close encounter with the
planet, which has Keplerian frequency $\omega$ and orbital radius $r_p$.
Both the planet and the test particle are initially on circular orbits.
In a single encounter, the particle's expected change in specific
angular momentum (Lin \& Papaloizou 1993) is
\begin{equation}
\delta h_e = -\frac{64}{243} \frac{G^2 M_p^2 r_o}{\omega^3
\Delta^5} \left [ 2K_0 \left ( \frac{2}{3} \right ) + K_1 \left
( \frac{2}{3} \right ) \right ]^2 ,
\label{deltah}
\end{equation}
where $M_p$ is the mass of the planet, $\Delta = |r_o - r_p|$, and $K_0$
and $K_1$ are modified Bessel functions (Goldreich \& Tremaine 1982).
The rate of change in the particle's specific angular momentum due to
repeated encounters with the planet is therefore
\begin{equation}
\dot{h} = \frac{\delta h_e}{\delta t} ,
\label{hdot}
\end{equation}
where the time between encounters is
\begin{equation}
\delta t = \frac{2 \pi}{|\Omega - \omega|} .
\label{deltat}
\end{equation}
The timescale over which the particle's specific angular momentum
changes by of order unity is simply
\begin{equation}
t = \frac{h}{\dot{h}} = \frac{r_o^2 \Omega}{\dot{h}} .
\label{tscaleeq}
\end{equation}

Figure \ref{timescale} shows the scattering timescale of disk particles
orbiting near two hypothetical planets, one with Jupiter's mass and the
other with Saturn's mass. Both planets are placed 10~AU from a Sun-like
star. Each planet's tidal radius, or Hill radius, is defined by
\begin{equation}
R_H = r_p \left ( \frac{M_p}{3M_*} \right )^{1/3} ,
\label{rhill}
\end{equation}
given a stellar mass of $M_*$. For test particles within three Hill
radii of the planet's orbit, the scattering timescale is $10^4$ years or
less. Tidal interactions with planets are therefore extremely efficient
at opening gaps in protoplanetary disks. Notably, Crida et al.\ (2006)
found that even in strongly accreting disks with kinematic viscosity of
order $10^{15} \: {\rm cm^2 \: s^{-1}}$ at 1~AU, corresponding to an
accretion rate of $\dot{M}_* \approx 5 \times 10^{-7} M_{\odot} \; {\rm
yr}^{-1}$ (Dodson-Robinson et al.\ 2009a), a Jupiter-mass planet ($1
M_J$) can open a gap in only 1000 orbits. Encouragingly, planets can
open gaps even in disks where strong accretion would readily fill in
gaps created by photoevaporative winds.

There is, however, a limited scope to the tidal influence of the planet.
As we see from Figure \ref{timescale}, the scattering timescale for
particles located four Hill radii or more from the planet approaches the
ages of transitional disk host stars. Setting a high scattering
timescale limit of $5 \times 10^5$~years, which is one-half to
one-quarter the estimated stellar age in the Taurus-Auriga association
\citep{palla00} and five times higher than the inferred transitional
disk lifetime/clearing time \citep{furlan09}, we find that the gap
opened by a planet may not exceed $\sim 5$~Hill radii.  And, since the
Hill radius scales as $M_p^{1/3}$, simply increasing the planet mass
does not provide a commensurate widening of the gap. Even for a $5 M_J$
planet at 10~AU, the Hill radius is only 1.2~AU, which allows a gap to
open from 4~AU--16~AU. This gap is simply not wide enough to account for
the observed hole sizes of any of the transitional disks listed in Table
\ref{table:stats}.

Given the sharp increase in scattering timescale as a function of
distance from the planet (Figure \ref{timescale}), our maximum gap width
is not highly sensitive to our assumed gap-opening timescale.
Our order-of-magnitude estimate of maximum gap width agrees with the
numerical work of Crida et al.\ (2006). Using analytical and numerical
calculations of the balance between torques from the planet and viscous
stresses, they found that a Jupiter-mass planet in a disk with aspect
ratio $H/r = 0.05$ opened a gap with a {\it full} radius of $5.5 \; R_H$
between the planet orbit and the primordial, unperturbed disk. The width
of the optically thin gap is, of course, much smaller than the full
center-to-wall width.  Our calculation of $5 \; R_H$ as the maximum
gap size is actually rather generous.

\subsection{Interior disk clearing by viscous accretion}
\label{accretion}

Our calculations demonstrate that a single planet cannot clear a gap
wider than $\sim 5$ tidal radii over typical T-Tauri star ages. A
Jupiter-mass planet at 10~AU could therefore open a gap extending from
6.5--13.5~AU. However, the star is accreting: could the accretion flow
drain the disk interior to the planet's orbit (which we hereby label the
``interior disk'') within the transitional disk lifetime and/or clearing
time?  For a disk with steady-state mass transfer, the diffusion
timescale is
\begin{equation}
t_{\rm diff} = \frac{M(r_g)}{\dot{M_*}},
\label{difftime}
\end{equation}
where $\dot{M_*}$ is the star accretion rate and $M(r_g)$ is the interior
disk mass. A derivation of Equation \ref{difftime} is in Appendix
\ref{appendix1}.  For narrow annuli of material located near the star,
so that $r_g \la 2$~AU, clearing of the interior disk by viscous
accretion seems possible.
Consider a disk with a surface density profile of
\begin{equation}
\Sigma(r) = \left ( 808 \; {\rm g \; cm}^{-2} \right ) \left (
\frac{1 \; {\rm AU}}{r} \right )
\label{GMAurdens}
\end{equation}
and an inner edge due to magnetospheric truncation at 0.1~AU. This
is the pre-clearing density profile of the GM~Aur disk constructed
by Hughes et al.\ (2009), who inferred a total disk mass of $0.16
M_{\odot}$ and an outer radius of 300~AU from millimeter imaging. If the
star accretes at a rate of $10^{-8} M_{\odot} \; {\rm yr}^{-1}$, the diffusion
timescale for material in the inner 2~AU is $10^5$~years, similar to
transitional disk lifetimes/clearing times (Furlan et al.\ 2009).

However, what about holes that are 20~AU or larger? Once can place a
planet at most $5 R_H$ from the desired edge of the hole, or at about 14~AU for a
$3 M_J$ planet orbiting a Sun-like star. (Since $R_H$ scales linearly
with the planet's orbital radius, we open the widest gap if we place the
planet as far from the star as possible.) The inside edge of the
planet's gap is then at about 8~AU. For the inferred GM~Aur density
profile, the diffusion timescale for material at 8~AU is now $4 \times
10^5$~years, edging toward a significant fraction of star ages in
the Taurus, Ophiuchus and Coronet associations \citep{furlan09}. Furthermore, the diffusion
timescale is a really an e-folding time (Appendix \ref{appendix1}), not
a time in which complete clearing of interior disk material occurs.

To check how many e-foldings it takes to make our interior disk
optically thin at millimeter wavelengths, where the wide transitional
disk holes have been imaged (Brown et al.\ 2008, Hughes et al.\ 2009,
Andrews et al.\ 2011), we use the opacity law
\begin{equation}
\kappa_{\lambda} = 0.02 \left ( \frac{1 {\rm mm}}{\lambda}
\right )^{\beta} {\rm cm^2 g^{-1}},
\label{submmopacity}
\end{equation}
\citep{beckwith90}. In Equation \ref{submmopacity}, $\kappa_{\lambda}$
is the total mass opacity (gas + dust) at a given wavelength $\lambda$,
$\beta$ is an exponent near unity, and gas and micron and/or submicron
grains are mixed with a mass ratio of 100:1. Throughout this
article we will use the convention that $\kappa_{\lambda}$
refers to an opacity based on total mass, gas and dust together,
and $\kappa_{\lambda,{\mathrm dust}}$ gives an opacity
referenced to the mass of dust grains only.

The vertical optical depth of the disk is simply
\begin{equation}
\tau_{\lambda} = \kappa_{\lambda} \Sigma.
\label{opdepth}
\end{equation}
A surface density of only 20~g~cm$^{-2}$ of gas and dust mixed with a
100:1 mass ratio is sufficient to make a disk optically thick, $\tau =
1$, at $\lambda = 0.4$~mm. Diffusing the 100~g~cm$^{-2}$ initially at
8~AU in our reconstructed GM~Aur disk down to optically thin levels
would take {\it two} e-foldings, or $8 \times 10^5$~yr. Even for our
hypothetical gap with an inner edge at 2~AU, for which the
characteristic diffusion timescale is seemingly reasonable, three
e-folding times are required to get to optically thin levels, or $3
\times 10^5$~yr. If a planet is far enough from its host star for the
outer edge of its gap to be at 15-20~AU or more, there is a severe
timescale problem with clearing its interior disk to optically thin
levels by viscous accretion. As we explore in the next section, the
mid-IR limits on leftover surface density from the interior disk are
even stronger than the limits from millimeter observations.




\subsection{Interior disk clearing by other mechanisms}
\label{otherclearing}

Interior disk material could be cleared via the scenario proposed by
Varni\`{e}re et al.\ (2006), in which spiral waves excited by a single
planet drive enhanced angular momentum transport that clears the
interior disk within $\sim 0.16 t_{\rm diff}$. Using 2-d hydrodynamic
simulations of planet-disk interactions, Varni\`{e}re et al.\ show that
within 2,000 orbits, the planet's spiral wakes reduce the interior disk
surface density by about a factor of 10. To test the viability of spiral
wake-driven clearing, we once again set a target hole size of 20~AU to
match the GM~Aur disk and place a hypothetical planet at 14~AU, $5 R_H$
from the edge of the hole. Our hypothetical planet takes $10^5$ years to
complete 2,000 orbits and open its incomplete hole---so far the clearing
time matches the transitional disk lifetime predicted by Furlan et al.\
(2009).

But an axisymmetric factor of 10 surface density reduction is not nearly
enough to make an optically thin hole in the mid-infrared. Using a
micron/submicron grain opacity of $\kappa_{\lambda, {\mathrm dust}} =
500$~cm$^2$~g$^{-1}$ at 13~$\mu$m \citep{ossenkopf92}, a gas/small grain
mass ratio of 100 and Equation \ref{opdepth}, we see that the total
(gas+dust) surface density required for a disk to be vertically
optically thin at 13~$\mu$m---the spectral index that usually diagnoses
flux deficits in transitional disk SEDs (Furlan et al.\ 2006)---is only
0.2~g~cm$^{-2}$.  For spiral waves to reduce the surface density from
100~g~cm$^{-2}$ to 10~g~cm$^{-2}$ at our 8~AU interior disk edge still
leaves 50 times too much material to have an optically thin disk at
13~$\mu$m.  Single planets and their spiral wakes therefore do not clear
enough material from the interior disk to explain observed transitional
disk SEDs.  However, our submicron grain opacity value is calculated
assuming the grains have maximum sizes of 1~$\mu$m or less. It is
possible that the combination of a single planet, wake-driven hole
clearing, and locking away 98\% of the solid mass in pebbles,
planetesimals and/or planet cores could be consistent with observed
transitional disk fluxes.

We have also considered whether photoevaporation or the
magnetorotational instability (``MRI'', Chiang \& Murray-Clay 2007) could
clear the interior disk material.  As we discuss in detail in \S
\ref{photoevaporation}, phoevaporative clearing is only efficient at
radii $\gtrsim3$ AU.  In such a scenario, the interior disk still clears
via viscous accretion, and is subject to the time constraints we have
just discussed. The MRI scenario proposed by \citet{chiang07} also
cannot clear the interior disk material.  If the inner disk is {\it
already} cleared of most small dust grains, X-rays can ionize material
at the edge of the hole and produce inside-out hole widening. The
accretion rate is then found to scale roughly as the square of the hole
size.  Inside-out MRI clearing is therefore not effective at {\it
initiating} the clearing of material from the interior disk (ie, at
small disk radii).

Transitional disks such as GM~Aur and LkH$\alpha$~330, with inner hole
radii of 20~AU and 50~AU, respectively (Hughes et al.\ 2009, Brown et
al.\ 2007), cannot be explained by a single orbiting planet.  Even if
one takes into account accretion or spiral wave-induced clearing of the
material left inside the planet's gap, the inner hole size that can be
cleared by a single planet in of order $10^5$~years is limited to a few
AU. However, multiple planets orbiting the same star would open adjacent
gaps which, combined, could form a large inner hole.  In the next
section, we use numerical simulations to demonstrate how the
``overlapping gap effect'' can explain even wide inner holes in
accreting disks.

\section{Dynamical clearing by multiple planets}
\label{multiplanets}

Our working picture of a transitional disk is an optically thick outer
disk surrounding a wide inner hole that produces little near- and
mid-infrared flux. The star must also be able to accrete material
through the hole at a high rate, as much as $10^{-8}
M_{\odot}$~yr$^{-1}$. Even though a single planet can only open a narrow
gap, a system of multiple planets could produce a wide hole if the
planet orbits were close enough together to create overlapping gaps.
Because the GM~Aur system is one of the best-characterized transitional
disks known, we use its overall mass, hole size and accretion rate (Table
\ref{table:stats}) as guidelines for our study of multiple planets in
transitional disks.

We describe our hydrodynamic simulations of hole opening by interacting
planets in \S \ref{hydro}.  Then, in \S \ref{holestructure}, we show that the
non-axisymmetric flow patterns produced by planetary systems are the best possible explanation 
for the high accretion rates of transitional-disk host stars.



\subsection{Hydrodynamic simulations}
\label{hydro}

Our planet-disk simulations were conducted using the publicly available
version of FARGO, a polar, grid-based 2-d hydrodynamic code designed
specifically for planet-disk interactions written by Frederic Masset
(Masset 2000). Like all polar hydrodynamic codes built on the
operator-splitting technique, FARGO uses a sequence of three
steps---source step, radial transport step, and azimuthal transport
step---to evaluate the transport of any hydrodynamic quantity.  The
FARGO transport algorithm differs from standard codes in that the azimuthal advection is
carried out in a reference frame that rotates at the local Keplerian
speed, so that hydrodynamic quantities are advected using the residual
velocity, $v_{\phi} - \bar{v}_{\phi}$, instead of the inertial azimuthal
velocity. A subsequent step corrects for the azimuthal shifts in the
field introduced by subtracting out the Keplerian speed. The advantage
of the FARGO transport algorithm is that timesteps are not
Courant-limited by the short dynamical time at the inner disk boundary.
FARGO produces up to a tenfold speedup in 2-d hydrodynamic calculations
over a traditional advection procedure, making simulations that cover a
significant fraction of the star age tractable.

In order to test whether a stable multiplanet system could reside inside
a transitional disk hole, our simulations include both migration-causing
torques from the disk and dynamical interactions between the planets.
Using the 20~AU hole size of GM~Aur's disk and our single-planet
gap size of $5 R_H$ as a guide, we set up a disk with three planets
located about six mutual Hill radii apart. Chatterjee et al.\ (2008)
defined the mutual Hill radius in planet encounters as
\begin{equation}
R_{Hm} = \left ( \frac{M_1 + M_2}{3M_*} \right )^{1/3} \left ( \frac{a_1 +
a_2}{2} \right ).
\label{mutualhill}
\end{equation}
Tightly-packed systems with planet orbits spaced less than $4 R_{Hm}$
apart strongly scatter on 1--10~Myr timescales and are not likely to
survive intact (Chatterjee et al.\ 2008, Scharf \& Menou 2009).  For
planet orbits $6 R_{Hm}$ apart, the outer $4 R_H$ boundary of the
innermost planet, measured in {\it individual} Hill radii, coincides
with the inner $4 R_H$ boundary of the next planet in the system. Our
orbital spacing therefore favors the creation of overlapping gaps, which
can reach $5 R_H$, yet maximizes the likelihood of long-term dynamical
stability. To recreate the 20~AU hole in the GM~Aur disk, we place three
planets, each of three Jupiter masses, at distances of 14.3~AU, 6.3~AU
and 2.7~AU. Our planetary configuration would also provide a reasonable
match for the holes of DM~Tau or DoAr~44 (19~AU and 30~AU, respectively; Andrews et al.\ 2011).

The disk has a constant aspect ratio of $H/r = 0.05$, where $H$ is the
pressure scale height. We place the outer boundary at 38~AU and the
inner boundary at 0.72~AU. Our evenly spaced grid contains 220 radial
zones and 384 azimuthal zones. The inner boundary is open so that mass
can leave the grid, simulating accretion onto the star. We also allow
mass to enter the grid at the outer edge of the disk, mimicking a
steady-state mass transfer profile.  The initial surface density profile
of the disk is the same as in Equation \ref{GMAurdens} (\S
\ref{oneplanet}).  If we extrapolate the surface density profile out to
300~AU, the observed size of the GM~Aur disk (Hughes et al.\ 2009), the
overall disk mass would be $0.16 M_{\odot}$.

Accretion through the disk is parameterized using the
$\alpha$-viscosity, where the turbulent diffusion coefficient is
\begin{equation}
\nu = \alpha c_s H.
\label{alpha}
\end{equation}
In Equation \ref{alpha}, $c_s$ is the sound speed and $\alpha$ is the
efficiency with which gravitational energy is extracted from the shear
flow and transformed into turbulence.  Our choice of $\alpha$ was
motivated by studies of turbulence caused by the magnetorotational
instability or MRI, the leading contender for the viscosity source in
protostellar disks (Balbus \& Hawley 1991, Balbus et al.\ 1996).
Turner \& Sano (2008) find that $\alpha \le 10^{-4}$ inside the
optically thick, nonionized ``dead zone'' within two pressure scale
heights of the disk midplane. We adopt a turbulent efficiency of $\alpha
= 0.002$, consistent with mass-weighted estimates of local accretion
rates with turbulence driven by the MRI (Turner \& Sano 2008, Turner \&
Drake 2009). All parameters in the FARGO simulation are summarized in
Table \ref{table:fargopars}.

Like Zhu et al.\ (2011), we allow FARGO to apply only the Type II
migration torques.  In Type I migration, which occurs before the planet
opens a gap, the disk's outward pressure gradient amplifies the torques
from the planet's wake at the outer Lindblad resonances, forcing the
planet to move inward at an extremely high rate (Ward 1997, Tanaka et
al.\ 2002). In the limit of low thermal diffusivity, the nonlinear
torque caused by the ``horseshoe'' of gas and dust corotating with the
planet saturates and the only net torque on the planet comes from the
Lindblad resonances (Ward 1991). Yet a new series of investigations of
planet migration in viscous disks with nonzero thermal diffusivity
demonstrates that the corotation torque may halt inward migration or
even drive outward migration (Paardekooper \& Mellema 2006, Kley \&
Crida 2008, Kley et al.\ 2009, Paardekooper et al.\ 2010, Paardekooper
et al.\ 2011). Since FARGO does not treat heat or momentum diffusion,
the migration rates calculated in the Type I limit---either no gap or an
incomplete gap---may not accurately describe the motion of a planet in a
physically realistic disk. To prevent improper computation of Type I
migration torques, we first follow the gap-opening process for 200
orbits of the outermost planet (12,000 years) with the migration feature
of the FARGO code turned off. We then ``release'' the planets,
turning on the migration torques and the N-body integrator to
track planet-planet interactions.

Unlike Zhu et al.\ (2011), we do not include accretion onto the planets,
which can be parameterized in FARGO by specifying the percentage of mass
entering the planet's Hill radius that the planet can accrete, $p$. Over
the course of both the initial 12,000-year gap-opening epoch in our
simulations and the $> 10^5$-year length of the hole-opening phase
(Furlan et al.\ 2009), a young gas giant shrinks by at least two orders
of magnitude, dramatically decreasing its accretion efficiency.  A
physically realistic estimate of $p$ would therefore require both
an adaptive mesh to model the circumplanetary disk in detail and robust
knowledge of the interior structure and evolutionary stage of each
planet.  Because the $3 M_J$ planets we have placed in our model
transitional disk should be much smaller than their Hill spheres for
most of the simulation, we have assumed that accretion onto the planets
clears a negligible amount of disk material. In Appendix \ref{appendix2}
we review the evolution of a young gas giant's interior structure in
more detail and discuss the effects of these changes on the
planet-growth process.

\subsection{Hole structure and appearance}
\label{holestructure}

Figure \ref{GMAurhydro} shows the surface density of gas and dust as a
function of time in our simulation. A wide hole created by three
overlapping individual gaps opens within about 300 orbits of the
outermost planet, or $1.8 \times 10^4$~years. The color scale in the
topmost row of Figure \ref{GMAurhydro} is designed to mimic sub-mm
opacities. Based on the opacity law in Equation \ref{submmopacity}, disk
gas and dust mixed with a 100:1 mass ratio with surface density $\sim
20$~g~cm$^{-2}$ is optically thick ($\tau > 1$) at 400~$\mu$m. All parts
of the disk rendered in white are optically thin in the sub-mm.  After
400 orbits of the outermost planet, or $2.4 \times 10^4$~years, the
optically thin sub-mm hole extends to about 17~AU. At $\lambda = 1$~mm,
the hole extends to about 20~AU. The appearance of a wide hole
consistent with sub-mm observations is not merely a result of having
assumed a low-density disk: our initial surface densities are entirely
based on the mass of the GM~Aur disk inferred from millimeter
observations (Hughes et al.\ 2009), which is quite high at $0.16
M_{\odot}$. Interestingly, Salyk et al.\ (2009) note that transitional
disks with measurable accretion tend to have higher inferred masses than
``classical'' disks without holes, which supports planet formation as a
likely explanation for observed hole structures (although the low-mass
transition disk fraction is not yet well known; Andrews et al.\ 2011).

Furthermore, all transitional disks have mid-IR flux deficits indicating
their inner holes are optically thin from 10-25~$\mu$m. On the bottom
row of Figure \ref{GMAurhydro}, we stretch the color scale to mimic
13~$\mu$m optical depth. At 13~$\mu$m, micron and submicron grains
produce an opacity of $\kappa_{\lambda,{\mathrm dust}} \sim
500$~cm$^2$~g$^{-1}$  \citep{ossenkopf92} that is not very sensitive to size distribution as
long as the maximum grain size is under 10~$\mu$m.
The $\tau = 1$ threshold is then at a dust-only surface density of
0.002~g~cm$^{-2}$ of micron and submicron grains.  Assuming a gas/solid
mass ratio of 100 and making the assumption that 100\% of the disk's
solids are in micron and submicron grains (as opposed to pebbles,
planetesimals or planet cores), the gas and dust together must have a
surface density $\Sigma \la 0.2$~g~cm$^{-2}$ to be vertically optically
thin at 13~$\mu$m. Again, all optically thin regions of the disk in the
small-grain approximation are rendered in white. Globally, the inner
hole has a filling factor of $f \approx 0.5$ and is overall optically
thin.

Where a significant portion of the grains have formed pebbles,
planetesimals or planets, the optically thin threshold surface density
of gas and solids mixed with a 100:1 mass ratio is higher than
0.2~g~cm$^{-2}$. In Figure \ref{highgassolid} we once again render the
400-$\mu$m and 13-$\mu$m optical depths calculated from orbits 197 and
395 of our hydrodynamical simulation, accounting for the fact that a
significant fraction of the available solid mass must be locked away in
giant planet cores, residual planetesimals, or planetesimal precursor
particles. Our fiducial opacity of $\kappa_{\lambda, {\mathrm
dust}} = 500$~cm$^2$~g$^{-1}$ for dust at
13~$\mu$m is valid to about 20\% for any single grain size or size
distribution as long as the maximum grain size present is of order
1~$\mu$m. Yet Mie scattering calculations by \citet{ossenkopf92} show
that for the MRN grain size distribution,
\begin{equation}
\frac{d \log n(a)}{d \log a} = -3.5,
\label{mrn}
\end{equation}
where $n$ is the number of grains and $a$ is the grain radius
\citep{mathis77}, significant 13-$\mu$m opacity reductions begin when
the maximum grain size is extended to 10~$\mu$m. Extending the MRN
distribution to to millimeter, ``pebble'' sizes produces a factor-of-10
opacity reduction over our fiducial value of
$\kappa_{\lambda,{\mathrm dust}}$. In Figure \ref{highgassolid}
we assume that 80\% of solids have either grown to millimeter sizes or
become incorporated in planetesimals/planets and the remaining 20\% of
the solids are micron and submicron grains which provide the observed
flux. The optically thin threshold surface density of gas and
solids mixed with a 100:1 mass ratio is
then $\Sigma \la 1$~g~cm$^{-2}$ at 13~$\mu$m and $\Sigma \la
100$~g~cm$^{-2}$ at 400~$\mu$m. At 13~$\mu$m, almost 90\% of the inner
hole's surface area contains only optically thin dust and the optically
thick tidal streams cover only 10\% of the hole, $f \approx
0.1$.


In Figures \ref{GMAurhydro} and \ref{highgassolid}, we see the detailed
structure of each planet's tidal tail, as well as small pileups of
material between the planets. Far from being problematic sources of
excess flux, these pileups and tails provide a critically important
feature of our model: the ability to transport mass through an
apparently optically thin hole. The apparent contradiction between
optically thin holes and high stellar accretion rates comes from the
assumption that gas and grains are evenly and axisymmetrically
distributed within the hole (Espaillat et al.\ 2008, Espaillat et al.\
2010, Zhu et al.\ 2011). For uniformly distributed grains and a
gas/solid mass ratio of 100, our 0.002~g~cm$^{-2}$ of dust, $\tau = 1$
threshold would allow less than eight {\it lunar} masses of dust in the
hole, or $0.1 M_{\oplus}$. If 80\% of the solid mass is in pebbles,
planetesimals or planet cores and only the remaining 20\% provides
13~$\mu$m opacity, the total mass of solids inside the hole could still
not exceed $0.5 M_{\oplus}$.

By contrast, after 24,000~years of evolution, our simulated disk's inner
hole contains about $0.95 M_{\oplus}$ of solids---yet this material
covers between 10\% and 50\% of the surface area of the hole, depending
on the assumed gas/small grain mass ratio.  In our fargo simulation, the
inner hole contains about Saturn's mass ($95 M_{\oplus}$) of gas and
dust together. We have therefore used the non-uniform, non-axisymmetric
property of flow patterns modulated by giant planets to at least double
the mass of small grains consistent with an apparently optically thin
inner hole.





Another new piece of observational evidence supports the idea that gas
in the inner holes of transitional disks is confined by planets.  Using
rotation diagrams of the CO fundamental rovibrational ladder, Salyk et
al.\ (2011) found that simple emission models are consistent with
systematically lower surface areas for transitional disks than for
gap-free protoplanetary disks.  The tidal barriers of planets in the
inner holes of transitional disks would confine hot gas to tidal tails
or narrow annuli between planets and therefore lead to a lower emitting
area for hot gas in transitional disks.

One issue highlighted by Zhu et al.\ (2011) was their difficulty keeping
the planetary system stable. They found that migration drives the
planets into 2:1 resonances, which leads to strong scattering. The high
migration rate in their simulations is due to their choice of $\alpha$
in the standard $\alpha$-viscosity parameterization: Zhu et al.\ used
$\alpha = 0.01$, where we use $\alpha = 0.002$, motivated by simulations
of MHD turbulence (\S \ref{hydro}). Type II migration operates on the
disk's viscous timescale, which is shorter for higher values of
$\alpha$. The planets in our simulations do begin to migrate after they
are released, particularly the outermost planet, which moves from
14.3~AU to 10~AU over the first $5 \times 10^4$ years of the simulation.
Yet our orbit integration shows that planetary system remains stable for
more than $10^5$~years. Although physical understanding of turbulent
viscosity generation in disks is still limited, it is possible to
construct a physically realistic transitional disk in which migration
does not destabilize the planetary system.

\section{Model transitional disk SEDs}
\label{modelseds}

In this section we consider whether the multiplanet dynamical model
shown in Figure \ref{GMAurhydro} is consistent with observed photometry
of transitional disks.  Our goal is to discuss the degree of
compatibility between our model and observed fluxes; an exact fit of SED
model to data is beyond the scope of this work, and would be hampered by
significant limitations.  Among the limitations is the fact the
dynamical model predicts only the gas distribution, while most
observables are based on emission from small dust grains. Because such
grains would be affected by pressure gradients set up by the complex gas
structure, and because the expected gas/small-dust-grain ratio is
essentially unknown, the translation from gas distribution to an
emission spectrum is not trivial.

Nevertheless, we make some simple assumptions here and produce models
designed to give us a qualitative understanding of the compatibility
between the multi-planet simulation and transitional disk observations.
In particular, we believe that the most stringent requirement is that
mid-IR fluxes be as low as observed, as most other features of the disk
SED can be fit by tweaking any of the numerous (and degenerate)
parameters that enter sophisticated disk models.  We begin by comparing
our model to photometric data of the GM~Aur system. However, we find
that our model is in some ways more naturally consistent with
pre-transitional disks.  We therefore also discuss the pre-transitional disk DoAr~44.

Our simple model consists of three components: an optically thin inner
disk from $R_\mathrm{in}$ to 0.7 AU, an optically thick component from
0.7-13 AU with a constant filling factor $f$ meant to represent the
dynamical model described in \S \ref{multiplanets}, and an optically
thick outer disk from 13 AU to $R_\mathrm{out}$ (see Figure
\ref{sed_plot}).  Model SEDs are created using the Monte Carlo radiative
transfer code RADMC \citep{dullemond04}, again without vertical
structure integration.  Stellar parameters and disk inclination are set
to known values \citep{hughes09, andrews11}.  We use a standard Milky
Way dust composition of 20\% carbon and 80\% small (sub-$\mu$m)
circumstellar silicates \citep{weingartner01, li01, draine07}.
$R_\mathrm{in}$ was set at the measured dust inner radius, if known, or
at roughly the dust sublimation radius, and the mass and radial extent
of optically thin dust were adjusted to roughly match observations.  To
simulate an outer disk that is not directly irradiated by stellar flux
but instead mostly shadowed by tidal streams, we took the difference
between a model from $R_\mathrm{in}$ to $R_\mathrm{out}$ and another
model from $R_\mathrm{in}$ to 13~AU---although, as is apparent in Figure
\ref{sed_plot}, the observations do seem to require partial illumination
to `puff up' the hole's outer edge \citep[e.g.][]{andrews11, hughes09}.
However, the flux from the outer disk is not the focus of this work, and
we have made no effort to fully explore parameter space for these
components.

Instead, we are concerned with the near- and mid-IR flux contribution
from the confined tidal streams in our dynamical model.  We begin by
discussing the mid-IR fluxes, which originate from the $\sim$ few AU
region of the disk.  The left-hand side of Figure \ref{sed_plot} shows
the azimuthally averaged dust surface density for our dynamical model
after 400 orbits of the outermost planet.  If one assumes the mass is
spread evenly over all azimuths the surface density is almost everywhere
optically thick at 13 $\mu$m (assuming an opacity of
$\kappa_{\lambda,{\mathrm dust}} = 500$~cm$^2$ g$^{-1}$ for dust
grains). Not surprisingly, since transitional disks by definition have
very low 13 $\mu$m fluxes, an optically thick disk model with $f = 1$
(red squares) is not consistent with observations.  In reality, however,
the density distribution predicted by the dynamical model is not
everywhere optically thick, as the mass is shepherded due to the
presence of the planets.  In the language of \S \ref{diskproperties},
the azimuthal filling factor, $f(r)$, (gray solid line), ranges from
$\sim$0.1-1 if the gas/solid mass ratio is 100 and all solids are in
micron and submicron grains. Extending the grain-size distribution up to
millimeter radii or locking away solids in planetesimals and planets
correspondingly lowers $f$ (see Figures \ref{GMAurhydro} and
\ref{highgassolid}).

Full 2-d modeling of the emergent flux from the distribution
predicted by our hydrodynamical simulation is beyond the scope of this
analysis; instead, we make the simple assumption that the flux scales
proportionally to some characteristic average value of $f$, and ask
whether the resulting fluxes are consistent with photometric data. We implicitly assume that optically thin
parts of the disk are completely devoid of grains, and so this can be considered
a `best-case' scenario. In reality, the optically thin regions would need to have an 
average $\tau$ lower than that derived from axisymmetric models.
On the other hand, we begin with a conservative assumption about the gas/small-dust-grain ratio, and
later discuss the implications of this choice.

The middle panel of Figure \ref{sed_plot} compares
model SEDs with photometry and IRS spectroscopy of GM~Aur.
The blue curve (triangles) in Figure \ref{sed_plot} shows the flux
produced if the inner hole has $f=0.5$, a representative value for our
dynamical model if the gas/dust mass ratio is 100. Although the flux is
significantly reduced from the optically thick case, fluxes are still
inconsistent with observations.  Instead, to match observations, $f$
must be significantly smaller, or $\sim0.03$ (black curve).  For
comparison, \citet{calvet05} estimate $\tau\sim0.002$ at 10 $\mu$m.  Note
that we have not made an effort to fit the silicate feature exactly;
however, one can add additional contrast by adding small regions
of optically thin dust in the inner disk, and/or by adjusting the grain
composition (and thus opacity).  Pre-transitional disks like DoAr 44 (right hand side of Figure
\ref{sed_plot}) are similarly inconsistent with the completely optically
thick model (red squares), as well as the $f=0.5$ case (blue triangles),
using standard assumptions.  However, since the mid-IR fluxes in this
case are significantly higher, the required reduction in flux is more
moderate --- close to $f=0.1$ (black curve).

Are the observations consistent with the dynamical model?  As discussed
in \S \ref{holestructure}, $f$ is essentially anti-proportional to the
fraction of solid mass locked in pebbles, planetesimals and planet
cores, so one way to match observations of a disk like GM~Aur is simply
to reduce the mass fraction of small grains in the inner disk, as also
suggested by \citet{zhu11}.  That much of the mass has been incorporated
into large grains is certainly a reasonable possibility, given the fact
that large planets have already formed.  It is important to keep
in mind that our simple SED models ignore the flux contribution
from areas where $\tau < 1$ locally, and so the reduction
in flux with changes in $f$ due to grain growth is optimistic.
Nevertheless, with a power-law size distribution of silicate grains with a minimum grain 
size of $a_\mathrm{min}=0.005 \mu$m,
setting $a_\mathrm{max}=1$ mm reduces 13 $\mu$m opacity by a factor of $\sim$10 relative
to a distribution composed entirely of small grains.  Therefore, even moderate 
grain growth can easily make these models a reasonable possibility.

As an alternative to flux reduction
by grain growth, one could geometrically constrain the small grains,
reducing their filling factor relative to that derived for the gas.  It is well known that orbiting
dust grains subject to a gas pressure gradient can migrate towards
pressure maxima \citep{weidenschilling77}, and so it is not unreasonable
to expect some additional shepherding of dust grains relative to the
gas. 

We now consider the near-IR fluxes.  One notable feature of our
dynamical model (see Figure \ref{GMAurhydro}) is the persistence of a
small annulus of optically thick material between the innermost planet
and the star.  The dynamical influence of the innermost planet does not
extend to these radii, or to radii smaller than 0.7 AU, which is the
inner boundary of the disk in our dynamical simulation. Thus, the very
inner disk is {\it not} dynamically cleared, and, if present,
contributes significant near-IR flux.  As discussed in \S
\ref{diskproperties}, it is possible to hide some optically thick, but
very geometrically thin, material in the GM Aur inner disk.  However,
near-IR fluxes are certainly inconsistent with the large optically thick
dust rim extending from 0.2 to 1.5 AU that the dynamical model leaves
behind.  Therefore, for disks like GM Aur, this inner disk material must
be cleared by another mechanism, such as viscous dissipation.  Our
system of planets was specifically designed to have a viscous clearing
timescale for inner disk material of $\sim10^5$ years or less, so this
possibility is within reason.

For so-called pre-transitional disks such as DoAr~44, however,
near-IR fluxes actually require a significant amount of material close
to the dust sublimation radius (Espaillat et al. 2010; see also Figure
\ref{sed_plot}).  This interior pileup, also found by \citet{rice06}, is
a natural consequence of multi-planet systems of the type
described here.  In addition, although pre-transitional disks are
essentially defined by the presence of this inner disk material, even
``traditional'' transitional disk sources typically have measurable
excess dust emission \citep[e.g.][]{eisner06, espaillat10} as well as
emission from hot CO gas \citep{rettig04, pontoppidan08, salyk09}
originating at $r < 1$ AU.  Thus, our model suggests that the range of
observed near-IR fluxes represents a range of stages of dissipation of
the material interior to the innermost planet.

\section{Other hole-opening mechanisms: Photoevaporation}
\label{photoevaporation}

In \S \ref{multiplanets} we showed that systems of multiple planets
provide a robust explanation for the observed characteristics of
transitional disks: high $\dot{M}$, wide holes that appear optically
thin, and confinement of hot gas to small emitting areas.  However,
another way of opening gaps in disks is to expose the disks to energetic
radiation. Here we discuss the viability of photoevaporation as a
hole-opening mechanism. We show that, although photoevaporation can be
responsible for transitional disk holes about 3--6~AU wide, it cannot
build the extremely wide holes that concern us here.

X-ray, extreme-UV and far-UV photons transfer kinetic energy to gas in
the disk, either directly or through the photoelectric effect, allowing
the gas to become unbound from the star.  Photoevaporative flows begin
at the graviational radius, where the kinetic energy of the radiatively
accelerated gas particle is equal to the escape speed from the star's
potential:
\begin{equation}
R_g = \frac{G M_*}{c_s^2}.
\label{gravradius}
\end{equation}
In Equation \ref{gravradius}, $c_s$ is the sound speed in the
heated upper layers of the disk.
The localization of the photoevaporative flow lends itself well to
opening gaps in disks, which may then widen into holes as the interior
disk drains on the viscous timescale. However, recall the
discussion of the viscous timescale in \S \ref{oneplanet}: in
order for accretion to drain away interior disk material within
the lifetime of the disk, the gap must be located within a few
AU of the star. Of X-ray, EUV and FUV photons, only the X-rays
have a gravitational radius that might meet our diffusion
requirements.

In X-ray photoevaporation, disk clearing begins at a characteristic
``clearing age'' of 2-4~Myr (Owen et al.\ 2010). A gap in the disk opens
at a characteristic radius of 3~AU and the outer and interior disks are
decoupled, so that the interior disk drains on the viscous timescale,
several $10^5$~years to become optically thin for the surface density
profile in Equation \ref{GMAurdens}. X-rays take approximately 10-20\%
of the clearing age, or $2 \times 10^5$--$8 \times 10^5$~years, to
deplete the material between 3 and 10~AU. Viscous accretion of material
inside 3~AU, to leave behind only optically thin dust, and outward
expansion of the gap therefore occur on roughly the same timescale. Once
the interior disk has been accreted, the disk outside 10~AU is eroded by
the X-ray field hitting the hole's inner wall directly on a timescale of
order the clearing age, a few million years. Accreting transitional
disks with inner holes between 3 and 10~AU wide are therefore consistent
with the X-ray photoevaporation model. Indeed, Owen et al.\ (2010) found
that approximately 50\% of observed transition disks could be explained
by X-ray photoevaporation.

A closer look at the X-ray photoevaporation process, however, reveals
why X-ray driven winds cannot explain our test sample of transitional
disks. Once clearing begins to proceed from 10~AU outward via the
direct X-ray field, there is no longer any interior mass reservoir to
provide accretion onto the star.  Accreting disks with large inner holes
therefore cannot be explained by X-ray driven winds. For a 20~AU inner
hole, such as in the GM~Aur disk, Owen et al.\ (2010) find a probability
of less than $2 \times 10^{-4}$ that the star should still be accreting
(see their Figure 12). Furthermore, according to the Owen et al.\
models, one would have to wait at least 4~Myr to get a 20~AU
hole---2~Myr at minimum for the initial gap to open and another 2~Myr
for it to move out to 20~AU. Their models predict that transitional
disks with wide holes should be older than the average disk population,
whereas Furlan et al.\ (2009) and Merin et al.\ (2010) observe no
correlation between transitional disk fraction and cluster age.

We next consider photoevaporation driven by EUV radiation from either
the central star or nearby, massive stars. EUV continuum radiation
photoionizes the upper layers of a disk so that hydrogen recombination
creates a diffuse radiation field that steadily evaporates H~\Rmnum{2}
gas from the disk surface (Hollenbach et al.\ 1994, Clarke et al.\
2001). Once the inner disk has become optically thin to Lyman continuum
radiation, the direct UV field also drives a photoevaporative wind that
drains the outer disk (Alexander et al.\ 2006a). However, extreme UV
photoevaporation only drives disk dissipation when the viscous mass flow
through the disk becomes weaker than the photoevaporative flow,
estimated by Gorti \& Hollenbach (2009) as
\begin{equation}
\dot{M}_{EUV} \la 10^{-9.4} M_{\odot} \: {\rm yr}^{-1} .
\label{mdotuv}
\end{equation}
The stars hosting transitional disks in our test sample have higher
accretion rates (Table \ref{table:stats}).  Models of UV
photoevaporation therefore indicate that it cannot be responsible for
clearing the holes in our test sample.


Finally, we consider FUV-induced photoevaporation, for which the
gravitational radius is outside 30~AU (Matsuyama et al.\ 2003). Such a
large gravitational radius is appealing as a mechanism of opening wide
holes, but once more, we have a timescale problem with clearing the disk
interior to the gravitational radius. For our reconstructed GM~Aur
surface density profile (Eq.\ \ref{GMAurdens}), the single e-folding
diffusion timescale from 30~AU is 1.7~Myr--higher than the age of GM~Aur
and other transitional disks in Taurus. Furthermore, because the
FUV-driven wind outside the gravitational radius is so much more
efficient than viscous accretion inside the gravitational radius,
FUV-induced photoevaporation tends to truncate disks rather than opening
inner holes (Matsuyama et al.\ 2003).


Our final piece of evidence that photoevaporation is not responsible for
the holes in our sample of transitional disks (Table \ref{table:stats})
is its inability to create non-axisymmetric structure: photoevaporative
clearing proceeds evenly outward from the gravitational radius.
Photoevaporation therefore does not set up the confined accretion flow
necessary to move mass efficiently through an optically thin hole.
Multiple planets provide the only disk-clearing mechanism that
can simultaneously account for the wide inner holes, high accretion
rates, confinement of hot gas, and asymmetry observed in transitional
disks.


\section{Conclusions and testable predictions}
\label{conclusions}

While the initial discovery of transitional disks with wide holes were
based on near- and mid-IR flux deficits (often measured by the
13~$\mu$m/30~$\mu$m flux ratio; Furlan et al.\ 2006), more detailed
observational follow-up of a subset of these disks has revealed a number
of unusual characteristics, including:

\begin{enumerate}

\item Holes of 15~AU--70~AU or more that are too wide to have
been cleared by a single planet (Brown et al.\ 2007, Brown et
al.\ 2009, Hughes et al.\ 2010, Espaillat et al.\ 2010, Andrews
et al.\ 2011);

\item Stellar-mass companions orbiting within 4~AU of the central star
ruled out at the 99.7\% confidence level for three of our disks
in Table \ref{table:stats} (Pott et al.\ 2010);

\item Accretion onto the star at rates $10^{-9} \la \dot{M} \la 10^{-8}
\: M_{\odot} \: {\rm yr}^{-1}$ (White \& Ghez 2001, Johns-Krull \&
Gafford 2002, Andrews et al.\ 2011);


\item Complex radial mass distributions of dust and gas derived from
SED models, millimeter images, IR veiling, and IR interferometry
\citep{espaillat10, andrews11, akeson11};

\item Non-axisymmetric structure derived from millimeter images,
spectro-astrometry and sparse aperture masking \citep{brown09,
pontoppidan08, huelamo11}.

\end{enumerate}

In this article, we have shown that multiple planetary systems are the
only phenomena that can explain all of the transitional disk
characteristics listed above. The other possible disk-clearing
mechanisms---massive single planet, companion star and
photoevaporation---have timescale problems for wide holes and do not
provide ways to hide accreting mass in holes that appear optically thin.

The planetary system configuration in the multiplanet model must satisfy
three constraints. First, each planet must be a gas giant of $\sim 1 M_J$ or
more so that it can open a gap. Second, the planets must remain on
stable orbits for the lifetime and/or clearing time of the transitional
disk lest they disrupt the outer, optically thick disk outside the hole.
Finally, for sources with extremely low near-IR fluxes such as GM~Aur,
the innermost planet must be located within $\sim 2$~AU of the central
star so that the interior disk pileup can diffuse away within the
hole-clearing timescale. For the disks such as DoAr~44 with large holes
in millimeter images but higher near-IR fluxes, the constraint on the location of the innermost
planet is somewhat relaxed.

Central to our success at finding a single, unifying theory for the
wide-hole transitional disks is our break away from assuming
axisymmetric, radially uniform dust and gas distributions. In
illustrating how accreting mass can be hidden in apparently optically
thin holes (Figures 1, \ref{GMAurhydro} and \ref{sed_plot}), we are
using a geometrical definition of optical depth---the macroscopic
fraction of an emitting surface area that is covered with light sources.
Tidal tails and narrow annuli of dust and gas trapped between planets
are locally optically thick, but cover only a small fraction of the
inner hole. Overall, then, the inner hole can appear optically thin, but
still transport mass onto the star. Our hydrodynamic simulations suggest
that taking into account the possibility of a non-uniform dust and gas
distribution at least doubles the amount of dust that is consistent with
an optically thin inner hole.

The assumption of axisymmetrically distributed gas and dust is
problematic because it may hide the dynamical effects of 
planetary systems.  The existence of confined ``funnel flows''
raises the possibility that time variability of the star
accretion rate or excess near-IR flux may be a signature of
planetary influence. Accretion rate or near-IR variability would
occur on the orbital timescale of the innermost planet.
\citet{espaillat11} detected variability on 2-3 year timescales
in all but two of the 14 transitional and pre-transitional disks
in their study of the Taurus and Chamaeleon star-forming
regions.



The level of axisymmetry in inner disk flow patterns is an observational
test of our multiplanet theory of transitional disk origins.  As
discussed in \S \ref{diskproperties}, non-axisymmetric structures have
been discovered in a number of transitional disks \citep{brown08,
pontoppidan08, huelamo11}.  Embedded protoplanets are also expected to
produce asymmetries in emission line profiles \citep{regaly10}.
Furthermore, the finding of Salyk et al.\ (2011) that hot CO gas in
transitional disks has a different, possibly more confined, radial
profile as compared to classical disks should be investigated further
with disk models.  If confirmed, hot gas confinement strongly indicates
a dynamical origin for transitional disk holes rather than
radiation-driven clearing.

Finally, we offer a possible explanation, based on our
multiplanet theory, for the intriguing SR~21 disk which we
mentioned in \S \ref{diskproperties}. SR~21 has a gas disk
with a  7~AU inner hole but a dust disk with a 30~AU hole
\citep{pontoppidan08, brown09, andrews11}. If grain growth and
planetesimal formation in SR~21 have proceeded to the point
where very few primordial, small grains remain, any observed
dust would be second-generation, supplied by colliding
planetesimals. As we have seen in \S \ref{gapwidth}, giant planets are
extremely effective at scattering test particles away from their
orbits. A system of gas giants extending to $\sim 30$~AU would
deplete the region of planetesimals, choking off any chance of
producing second-generation dust. Gas, however, could still
accrete through the planetary system, explaining the smaller gas
truncation radius. The SR~21 system fits with our multiplanet
theory as long as the observed dust is primarily the product of
planetesimal collisions. 

As high angular-resolution astronomy advances, we urge observers to
focus planet-hunting campaigns on transitional disks. ALMA will provide
detailed kinematic and spatial information about the gas and dust
distributions in transitional disk holes. For the extremely wide, 70~AU
holes reported by Andrews et al.\ (2011), direct imaging with GPI may
yield pictures of young planets in formation. Finally, aperture-masking
techniques have the potential to reveal planetary-mass objects within $\sim 15$~AU of
their central stars \citep[e.g.][]{kraus09} and may uncover the hidden
planetary systems we suggest are inside the transitional disk holes.  In fact,
such a companion may have already been detected in the wide transitional
disk T~Cha \citep{huelamo11}.  Intriguingly, the only planets that can open gaps in disks are gas
giants of order Jupiter's mass. Transitional disks have excellent
discovery potential for young, still-forming planets, as planet searches
focusing on young stars are sensitive only to massive planets.

The discovery of extremely wide holes in transitional disks (Brown et
al.\ 2007, Andrews et al.\ 2011) points to a possible
gravitational-instability origin for the planets contained in the holes:
Dodson-Robinson et al.\ (2009b) demonstrated that gas giants with
semimajor axes $r_p \ga 35$~AU can only form by gravitational
instability. As a bonus, gravitational instability tends to produce
hotter, more massive planets than core accretion (Clarke 2009,
Stamatellos \& Whitworth 2009), exactly the types of planets best suited
to discovery by direct imaging and aperture masking.  In contrast,
radial-velocity observations of main-sequence stars, which have provided
almost all of the 500 and more exoplanets discovered to date, are
strongly biased toward planets formed by core accretion, which is
effective for shorter-period orbits. Short-period planets' inability to
open wide holes in disks could also explain why radial velocity surveys
of young stars with disks, which were designed to include large
fractions of transitional disks \citep[e.g.][]{prato08}, have yet to
detect a hot Jupiter.  In any case, studies of transitional disks will
likely provide insight into a newly confirmed and only partially
understood planet formation mechanism.

\acknowledgments
The authors would like to thank undergraduate intern Jeffrey Kerbow for
his diligent work in Summer 2010 and Joanna Brown, Sean Andrews and
Kevin Covey for insightful discussions about transitional disk
observations. We thank Kees Dullemond for making available his radiative
transfer code, RADMC, and its supporting documentation, and Frederic
Masset for releasing and documenting his 2-d hydrodynamic code, FARGO.
Support was provided by a University of Texas Dean's Fellowship awarded
to S.D.R., the Harlan J.\ Smith Postdoctoral Fellowship awarded to C.S.,
and the John W. Cox Endowment for excellence in astronomical research.
We also acknowledge computational support from the Texas Advanced
Computing Center.

\appendix

\section{Protostellar disk diffusion timescale}
\label{appendix1}

To calculate the rate at which disk material accretes onto the
star, we consider the surface density diffusion equation
(Pringle 1981):
\begin{equation}
\frac{\partial \Sigma}{\partial t} = \frac{3}{r}
\frac{\partial}{\partial r} \left [ r^{1/2}
\frac{\partial}{\partial r} \left ( \Sigma \nu r^{1/2} \right )
\right ] ,
\label{diffusion}
\end{equation}
in which $\Sigma$ is the surface density of disk material, $r$ is the
distance from the star, $t$ is time, and $\nu$ is the kinematic
viscosity of the disk material. Over a small region of the disk, one can
approximate the diffusion rate by assuming a constant viscosity so that
the diffusion equation can be separated into a time component and a
space component:
\begin{equation}
\Sigma(r,t) = R(r)T(t).
\label{sepvar}
\end{equation}
\citet{masset00} and \citet{masset03} also used disks with constant
kinematic viscosity in their analytical and numerical studies of planet
migration.
By taking the partial derivatives of the separated function
$\Sigma(r,t)$, we rewrite the diffusion equation as
\begin{equation}
\frac{1}{\nu} \frac{T'}{T} = \frac{9}{2r} \frac{R'}{R} + 3
\frac{R''}{R} = -\lambda,
\label{eigen}
\end{equation}
where $T'$ and $R'$ are partial derivatives over $t$ and $r$,
respectively, and $\lambda$ is the constant Eigenvalue. The solution
to the time-dependent portion of equation \ref{eigen} is
\begin{equation}
T(t) = \exp \left (-\lambda \nu t \right ) .
\label{Toft}
\end{equation}

To find an expression for $\lambda$ that is benchmarked against measured
T-Tauri star accretion rates and does not depend on assumptions about
viscosity, we examine the mass flow through the interior disk.  For an
interior disk annulus of finite size such as the one left behind by a
gap-clearing planet, one may approximate the total mass transfer through
the annulus, $\dot{M}$, as
\begin{equation}
-\dot{M} \approx \pi r_g^2 \dot{\Sigma}(r_g) ,
\label{sigmadotapprox}
\end{equation}
where $r_g$ is the inner edge of the gap created by the planet.  The
negative sign reflects the convention that material flowing inward
creates a positive accretion rate onto the star.  Substituting
$\dot{\Sigma}(r) = R(r) T'(t)$ into equation \ref{sigmadotapprox} and
solving for $\lambda$, we find
\begin{equation}
\lambda \approx \frac{\dot{M}}{\pi r_g^2 \nu \bar{\Sigma}},
\label{lambdasolved}
\end{equation}
where $\bar{\Sigma}$ is the average surface density in the diffusing
region $r < r_g$.

Using Equation \ref{lambdasolved}, we may write the
solution to the diffusion equation under the assumption of constant
viscosity as
\begin{equation}
\Sigma(r,t) \approx \Sigma(r,0) \exp \left (
\frac{-\dot{M}_0 t}{\pi r_g^2 \bar{\Sigma}} \right ) ,
\label{sigmasol}
\end{equation}
where $\dot{M}_0$ is the accretion rate at $t = 0$.  Here the diffusion
timescale we recover is similar to the canonical value $t_{\rm
diff} = r^2 / \nu$
(Lynden-Bell \& Pringle 1974), but in terms of the accretion rate:
\begin{equation}
t_{\rm diff} = \frac{\pi r_g^2 \bar{\Sigma}}{\dot{M}_0}.
\label{tdiff}
\end{equation}
The quantity $\pi r_g^2 \bar{\Sigma}$ is, of course, equivalent
to the disk mass $M(r_g)$ contained inside a circle of radius $r_g$, so
we rewrite Equation \ref{tdiff} to get Equation \ref{difftime}
in \S \ref{oneplanet}.

For disks with $\Sigma \propto r^{-1}$, as is usually inferred from
sub-mm observations, the diffusion timescale from Equation
\ref{difftime} is linearly proportional to $r_g$. A standard
$\alpha$-viscosity disk model with steady-state mass transfer has $\nu
\propto r$ so that the radial dependence of $t_{\rm diff}$ according to
the traditional $r^2 / \nu$ criterion is also linear. Even though we
have, for the purpose of clarity, based our derivation of Equation
\ref{difftime} on a disk with constant viscosity, the linear dependence
of $M / \dot{M_*}$ on radius applies to a range of commonly used disk
models.

\section{Accretion efficiency and planetary interior structure}
\label{appendix2}

Bodenheimer (1974) divides the evolution of a young gas giant into four
phases. The first is a transient hydrodynamic epoch, in which gas at the
planet's Hill radius free-falls toward the center of the planet: the
planet's rapid growth is driven by the runaway expansion of its Hill
sphere. The hydrodynamic epoch lasts about one free-fall time, $\sim
1,000$ years. For planets growing by core accretion, Phase I encompasses
the mass range of $\sim 10$--$500 M_{\oplus}$.  For planets formed by
disk fragmentation, the characteristic Phase I mass is much higher
(e.g.\ Clarke 2009). Planets in Phase I accrete 100\% of the material
inside their Hill spheres. Gap opening begins during Phase I at about
Saturn's mass, $95 M_{\oplus}$, and eventually halts the runaway growth
of the planet, which is why we do not start our simulations with
small planet cores and let them accrete: FARGO's accretion efficiency
parameter $p$ would suffer a dramatic drop during only the first 1,000
years of disk evolution.

At the onset of Phase II, the young planet develops an interior pressure
gradient that halts the free-fall of material at the Hill radius.  The
planet then stays in Phase II, quasistatic contraction on the
Kelvin-Helmholz timescale over about a factor of 20 in radius, for $\sim
10^5$ years. The planet's evolution during this period is similar to
that of a protostar on the Hayashi track. Accretion during Phase II is
limited by the disk's ability to deliver gas to the edge of the gap and
by the strength of the planet's tidal barrier (e.g. Dobbs-Dixon et al.\
2008). ``Leakage'' of gas across the tidal barrier is much more
efficient at the beginning of Phase II, when the planet nearly fills its
Hill sphere: at this stage, material from the disk can fall directly to
the planet's surface. As the planet contracts, the velocity of disk
material entering the Hill sphere increases relative to the planetary
surface, so that accreting gas will orbit the planet in a
circumplanetary disk (Papaloizou \& Nelson 2005).  Mass transfer onto
the planet is then regulated by the the diffusion timescale in the
circumplanetary disk as well as the mass transfer rate through the
protostellar disk and is extremely inefficient (Ward \& Canup 2010).

Once the planet's interior reaches 2000~K, molecular hydrogen begins to
dissociate, draining the thermal energy that maintains the quasistatic
pressure gradient. The resulting supersonic collapse, which starts Phase
III, takes only a few days and decreases the planet's radius by a factor
of more than 100, further limiting the planet's ability to accrete. We
see that a young planet's accretion efficiency changes dramatically as
it evolves.  During most of the transitional disk epoch, the planet will
be accreting through a circumplanetary disk whose mass transfer
efficiency depends on the ratio of protostellar/circumplanetary disk
viscosities and the size of the planet. The circumplanetary disk
viscosity depends in turn on the density and temperature of
circumplanetary material.

Since FARGO is an isothermal code that does not treat the planetary
interior structure, it much better suited for parameter studies of
gap-opening efficiency in the presence of planetary accretion than
physically realistic computations of planet growth profiles. Because the
accretion-limiting circumplanetary disk should be present and growing
during most of the hole-clearing epoch, we are justified in assuming
that mass transfer onto the planets is negligible throughout our FARGO
simulation.


\clearpage
\begin{figure}
\plotone{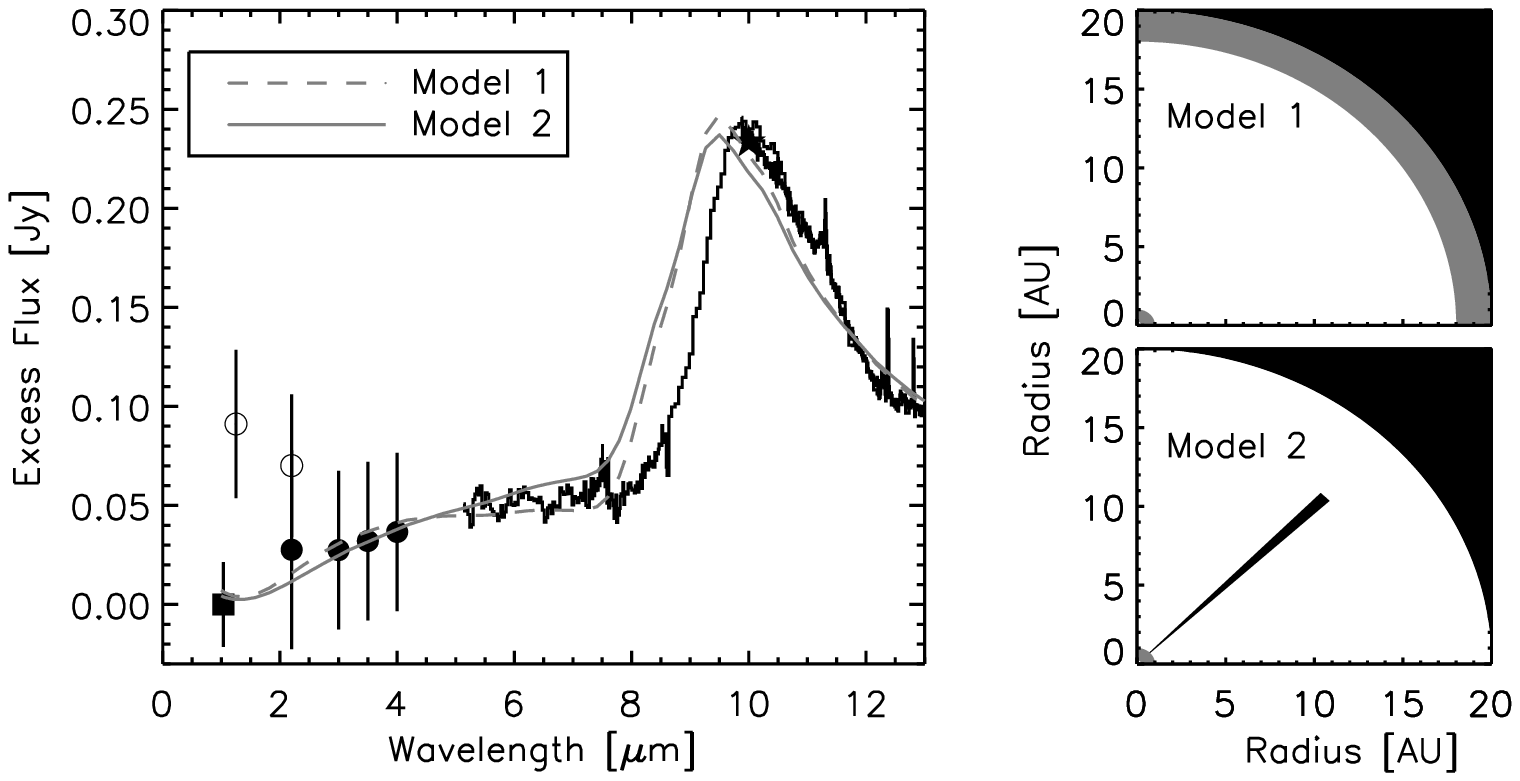}
\begin{minipage}{\textwidth}
\caption{In this figure we demonstrate some of the degeneracies inherent
to modeling of near- and mid-IR excess flux, especially once radial and
azimuthal asymmetry are permitted. The left panel shows near- and mid-IR
excess fluxes measured for GM Aur with two model SEDs based on
different mass distributions (square: Edwards et al. 2006; open
circles: Folha \& Emerson 1999
\protect\footnote[1]{These veiling values may be high due to a mismatch in photospheric template, or variability \citep{espaillat10}.}
; filled circles: Espaillat et al. 2010
\protect\footnote[2]{Last three points are estimated from Figure 5 in \citet{espaillat10} and are assigned the error bar associated with $r_K$.}
; star: Weaver \& Jones 1992; spectrum: Calvet et al. 2005,
Pontoppidan et al. 2010).
\protect\footnote[3]{We omit the M-band veiling values because of a likely template mismatch in their calculation \citep{salyk09}.}
Fluxes are de-reddening using $A_V=1.2$ \citep{espaillat10} and the reddening law of \citet{cardelli89}.
 The model mass distributions are shown schematically, and to scale, in the right hand panels, where gray represents optically thin regions, black represents
 optically thick regions, and white represents empty regions.  The outer disk contribution from R$>$20 AU is not shown, but
 primarily contributes at $\lambda>13 \mu m$. }
 \end{minipage}
\label{fig:veil_plot}
\end{figure}

\clearpage
\begin{figure}
\plotone{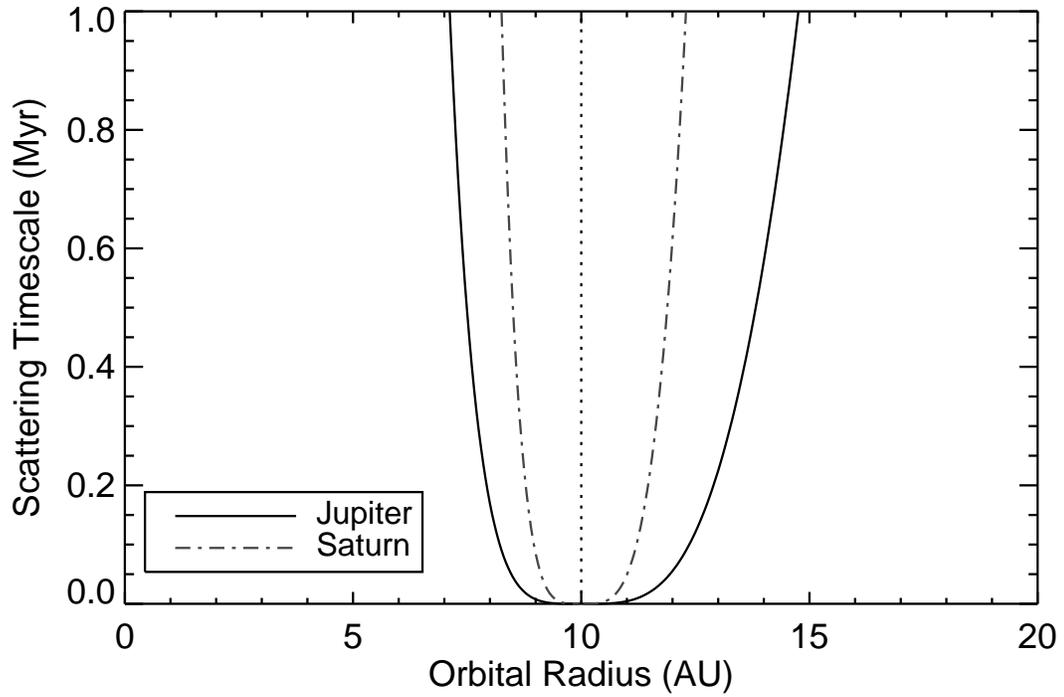}
\caption{The scattering timescale of particles initially on circular
orbits approaches the protostellar disk lifetime for particles more than
$\sim 5$~Hill radii from a planet. Here we plot scattering timescale as
a function of distance from the star in a disk containing a planet at
10~AU (vertical dotted line). The high scattering
timescales interior to the planet's orbit demonstrate that some other
disk dissipation mechanism must act along with planetary clearing to
clean out the inner holes of transitional disks.}
\label{timescale}
\end{figure}

\clearpage
\begin{figure}
\epsscale{0.9}
\plotone{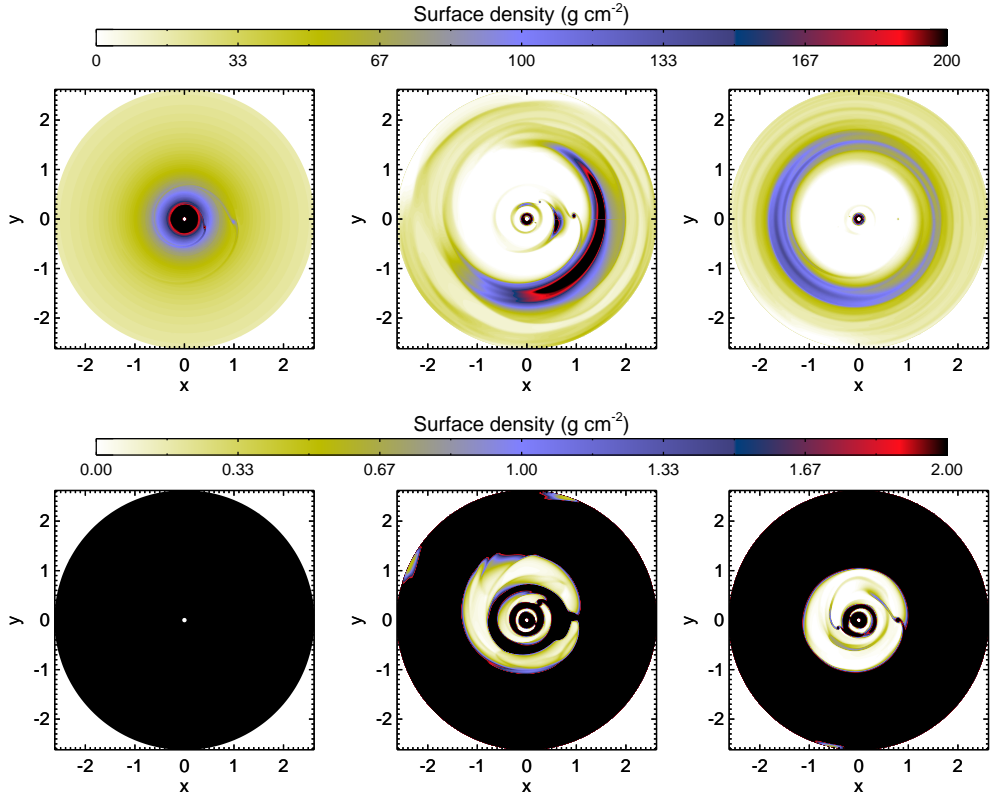}
\caption{{\bf Top Row}: At 20~AU wide, the hole in the GM~Aur disk is
too large to have been cleared by a single planet. Three medium-mass
planets, all $3 M_J$ with distances 14.3~AU, 6.3~AU and 2.7~AU from
GM~Aur, clear gaps that overlap and form a large hole over the course of
300 orbits of the outer planet ($2 \times 10^4$~years). Here we show
plots of the disk surface density from orbits 2, 197, and 395 of the
FARGO simulation, with the color scale chosen so white regions are
optically thin at millimeter wavelengths. The axes show position in the
corotating frame, where the outermost planet is at (x,y)~=~(1,0).  {\bf
Bottom Row}: Here we demonstrate that tidal streams can efficiently
transport mass through a hole that appears to be optically thin in the
near- and mid-infrared.  Simulation timepoints are the same as in the
top row, but the color scale is designed so that all white regions are
optically thin at 13~$\mu$m---assuming gas/solid mass ratio of 100, all
of the disk's solid mass is in micron or submicron grains, and
$\kappa_{\lambda, {\mathrm dust}} = 500$~cm$^2$~g$^{-1}$ for the grains
\citep{ossenkopf92}. Note how the non-axisymmetric, confined flow
pattern produced by the planets keeps most of the hole empty while
allowing mass transport through locally optically thick tidal streams.}
\label{GMAurhydro}
\end{figure}

\clearpage
\begin{figure}
\epsscale{0.6}
\plotone{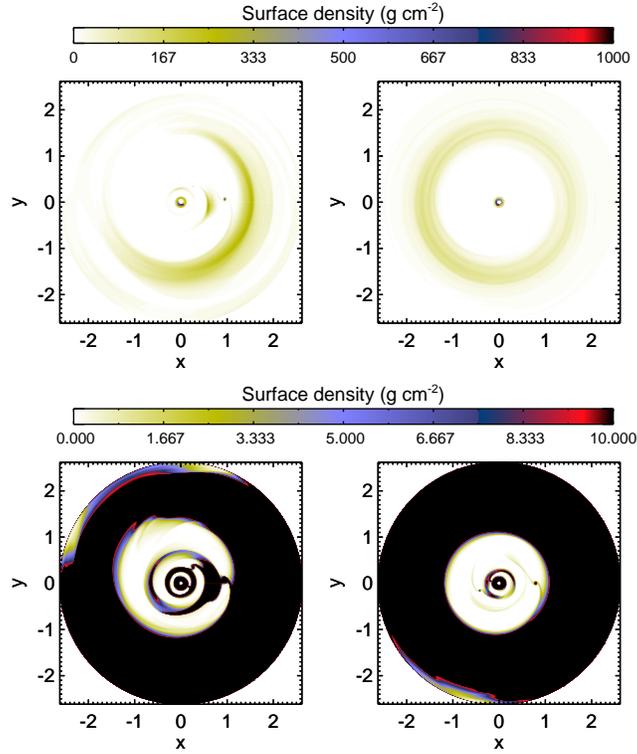}
\caption{{\bf Top Row}: Renderings of the 400~$\mu$m optical depth of
our model disk at orbits 197 and 395, here assuming that 80\% of the
disk's solid mass is in planetesimals and planet cores
that do not contribute to millimeter flux. As in
Figure \ref{GMAurhydro}, all white regions are optically thin. {\bf
Bottom Row}: The 13~$\mu$m optical depth of our model disk if 80\% of
the solids have formed pebbles, planetesimals or planets. After 400
orbits, locally
optically thick dust is confined to only about 10\% of the hole
surface area, yet the inner hole contains about Saturn's mass in
gas and solids. Here we assume an opacity of $\kappa_{\lambda} =
500$~cm$^2$g$^{-1}$ for micron/submicron grains. The total gas/solid
mass ratio is still 100.}
\label{highgassolid}
\end{figure}

\clearpage
\begin{figure}
\epsscale{1.0}
\plotone{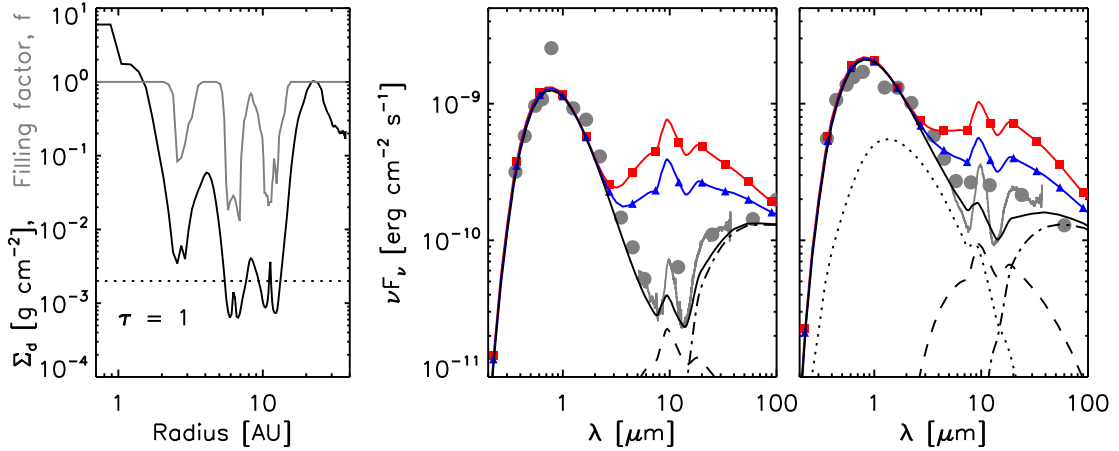}
\caption{{\bf Left}: While the azimuthally-averaged dust surface density
$\Sigma_d$ in our model transitional disk (black solid line) implies
optically thick dust everywhere except in two narrow annuli near the
outermost planet, taking into account azimuthal asymmetries yields a
filling factor, $f$, (gray solid line) less than unity over a substantial
portion of the inner disk. The dotted line denotes the optically thin
dust surface density cutoff at 13 $\mu$m. The filling factor
$f(r)$ was computed assuming a gas/small grain mass ratio of
100:1, but realistic gas/grain ratios in planet-forming disks
may be much higher (see \S \ref{holestructure} and Figure \ref{highgassolid}). {\bf Center:} A
comparison between the observed SED of GM Aur, shown in gray \citep{kenyon95,  weaver92, hartmann05, calvet05, pontoppidan10}, and fluxes produced using our reference multi-planet model. 
Models consist of an optically thin inner disk (dotted line, but below plot range for GM Aur), an optically thick inner disk
meant to represent our dynamical model (dashed line shows best-fit case), and an optically
thick outer disk (dot-dashed line).   From top to bottom, models have $f=1$ (red squares), $f=0.5$ (blue triangles) and $f=0.03$ (black). {\bf Right:} A comparison between the observed SED of DoAr 44 (references shown above plus Herbst et al.\ 1994) and our reference model.  From top to bottom, models have $f=1$ (red squares), $f=0.5$ (blue triangles) and $f=0.1$ (black). 
}
\label{sed_plot}
\end{figure}
\clearpage


\begin{deluxetable}{cccccc}
\tablecaption{Observed hole sizes and accretion rates of
transitional disks
\label{table:stats}}
\startdata
Name & hole size &  Ref & $\dot{M}$ & Ref \\
& [AU] & &[log(M$_\odot$ yr$^{-1}$)] && \\
\hline \\
DM Tau &19&\ref{An11} & -8.7 &\ref{Wh01} \\
DoAr 44 & 30&\ref{An11}&-8.0 &\ref{Es10}  \\
GM Aur &20&\ref{Hu09}&-8.0&\ref{Jo02}\\
HD 135344&39&\ref{Br09}&-8.3 &\ref{Ga06}\\
Lk H$\alpha$ 330 &68,47&\ref{An11}, \ref{Br09}&-8.8&\ref{Fe95}\tablenotemark{a}\\
MWC 758 & 73 & \ref{An11}&-8.0&\ref{An11} \\
SR 24 S& 29& \ref{An11} &-7.2 & \ref{Na06}\\
T Cha &15&\ref{Br07} &-8.4&\ref{Sc09}\\
UX Tau A & 25 &\ref{An11}& -8.0 & \ref{Es10} \\
\enddata
\tablerefs{
\usecounter{minirefcount}
\miniref{Andrews et al. 2011}{An11}
\miniref{Brown et al. 2007}{Br07}
\miniref{Brown et al. 2009}{Br09}
\miniref{Espaillat et al. 2010}{Es10}
\miniref{Fernandez et al. 1995}{Fe95}
\miniref{Garcia Lopez et al. 2006}{Ga06}
\miniref{Hughes et al. 2009}{Hu09}
\miniref{Johns-Krull \& Gafford 2002}{Jo02}
\miniref{Natta et al. 2006}{Na06}
\miniref{Schisano et al. 2009}{Sc09}
\miniref{White \& Ghez 2001}{Wh01}
}
\tablenotetext{a}{Estimated from full width of H$\alpha$ at 10\% level and Equation (1) of \citet{natta04}.}
\end{deluxetable}
\clearpage
\begin{deluxetable}{cll}
\tablecaption{Parameters of FARGO simulation
\label{table:fargopars} }
\tablehead{
\colhead{Parameter} & \colhead{Description} & \colhead{Value}
}
\startdata
\multicolumn{2}{l}{\bf Disk setup} \\
$r_{\rm in}$ & inner boundary & 0.72~AU \\
$r_{\rm out}$ & outer boundary & 38~AU \\
& grid zones & 220 radial, 384 azimuthal \\
& boundary conditions & open with outer mass
source\tablenotemark{a} \\
$H/r$ & aspect ratio & 0.05 \\
$\alpha$ & viscous efficiency & 0.002 \\
 & planet release time\tablenotemark{b} & 12,000~years \\
\multicolumn{2}{l}{\bf Planet setup} \\
$M_p$ & planet masses & $3 M_J$ \\
$r_p$ & initial orbit radii\tablenotemark{c} & 2.7~AU, 6.3~AU,
14.3~AU \\
& initial planet separation & $6 R_{Hm}$ \\
\enddata
\tablenotetext{a}{Open boundaries allow mass to move freely
through the grid and accrete onto the star. The outer mass
source allows material from beyond 38~AU, which we do not model,
to move into the grid assuming steady-state accretion.}
\tablenotetext{b}{To remove unphysical Type I migration torques
from our simulation, we hold the planets on fixed circular
orbits for 12,000~years (200 orbits of the outermost planet),
then turn on migration and fewbody interactions.}
\tablenotetext{c}{All planets were initially placed on circular
orbits.}
\end{deluxetable}

\end{document}